\documentclass[12pt,a4paper,english,amssymb,nofootinbib,superscriptaddress]{revtex4}
\usepackage{lmodern}
\usepackage{lmodern}
\usepackage[T1]{fontenc}
\usepackage[latin9]{inputenc}
\usepackage{color}
\usepackage{babel}
\usepackage{verbatim}
\usepackage{amsmath}
\usepackage{amssymb}
\usepackage{graphicx}
\usepackage{esint}
\usepackage[unicode=true,pdfusetitle,
 bookmarks=true,bookmarksnumbered=false,bookmarksopen=false,
 breaklinks=false,pdfborder={0 0 1},backref=false,colorlinks=true]
 {hyperref}
\hypersetup{
 linkcolor=blue,citecolor=blue}

\makeatletter

\pdfpageheight\paperheight
\pdfpagewidth\paperwidth

\@ifundefined{textcolor}{}
{%
 \definecolor{BLACK}{gray}{0}
 \definecolor{WHITE}{gray}{1}
 \definecolor{RED}{rgb}{1,0,0}
 \definecolor{GREEN}{rgb}{0,1,0}
 \definecolor{BLUE}{rgb}{0,0,1}
 \definecolor{CYAN}{cmyk}{1,0,0,0}
 \definecolor{MAGENTA}{cmyk}{0,1,0,0}
 \definecolor{YELLOW}{cmyk}{0,0,1,0}
}

\usepackage{babel}
\usepackage{babel}
\usepackage{babel}
\usepackage{babel}
\usepackage{babel}
\usepackage{babel}
\@ifundefined{definecolor}{color}{}
\usepackage{babel}

\usepackage{latexsym}\usepackage{bm}

\makeatother

\begin{document}

\title{From Smooth Curves to Universal Metrics}

\author{Metin Gürses}

\email{gurses@fen.bilkent.edu.tr}

\selectlanguage{english}%

\affiliation{{\small{}Department of Mathematics, Faculty of Sciences}\\
{\small{}Bilkent University, 06800 Ankara, Turkey}}

\author{Tahsin Ça\u{g}r\i{} \c{S}i\c{s}man}

\email{tahsin.c.sisman@gmail.com}

\selectlanguage{english}%

\affiliation{Department of Astronautical Engineering,\\
 University of Turkish Aeronautical Association, 06790 Ankara, Turkey}

\author{Bayram Tekin}

\email{btekin@metu.edu.tr}

\selectlanguage{english}%

\affiliation{Department of Physics,\\
 Middle East Technical University, 06800 Ankara, Turkey}

\date{\today}
\begin{abstract}
A special class of metrics, called universal metrics, solve \emph{all}
gravity theories defined by covariant field equations purely based
on the metric tensor. Since we currently lack the knowledge of what
the full of quantum-corrected field equations of gravity are at a
given microscopic length scale, these metrics are particularly important
in understanding quantum fields in curved backgrounds in a consistent
way. But, finding explicit universal metrics has been a hard problem
as there does not seem to be a procedure for it. In this work, we
overcome this difficulty and give a construction of universal metrics
of $d$-dimensional spacetime from curves constrained to live in a
$\left(d-1\right)$-dimensional Minkowski spacetime or a Euclidean
space. 
\end{abstract}
\maketitle

\section{Introduction}

\noindent There is a non-ignorable problem in high energy gravity:
we do not know the full field equations and the microscopic degrees
of freedom responsible for gravity. What we know is that Einstein's
theory is an effective one which will be modified with powers of curvature
and its derivatives, (most probably) in a diffeomorphism invariant
way, as long as the Riemannian spacetime model remains intact as a
valid description of gravity. At this stage, there is no compelling
reason to suspect that such a description ceases to make sense well
below the Planck scale. One might be deterred to say anything about
high energy gravity, in the absence of what the theory is, but the
situation is not that bleak as there are certain types of spacetimes
that solve \emph{any} metric-based equations. This approach to high
energy gravity is a remarkable one which started long ago \cite{Gibbons,Deser}
not exactly in this language, but developed \cite{guv,Horo,coley1}
over the years and culminated in a rather nice summary \cite{Coley},
where the notion of universal metrics with further refinements was
made, see also the more recent discussion in \cite{Universal-TypeIII-N,Universal-TypeII}.
Note that our definition of a universal metric is somewhat different
from the one defined in the previous literature: namely, for us, a
universal metric is a metric that solves all gravity theories defined
by covariant field equations purely based on the metric tensor. (We
shall not go into that distinction here and also not distinguish ``critical''
versus ``non-critical'' metrics, where the former extremize an action
while the latter solve a covariantly conserved field equation not
necessarily coming from an action.) 

The interest in the universal metrics is actually two-fold: these
are valuable on their own as they are solutions to putative low energy
quantum gravity at any order in the curvature. But, as importantly,
when one does quantum field theory at high energies, working about
these solutions will provide a better, self-consistent, picture as
gravity also plays a role. Of course, such metrics are hard to find,
as we are not given what the equations are. Therefore, one will be
hard-pressed to find, in the literature, examples of these metrics
save the examples given in the papers noted above. Perhaps, more important
is the fact that there is really no well-defined procedure of finding
these solutions except trial and error: namely, given a rather symmetric
metric, one can compute all possible curvature invariants and hope
that they vanish or at best they are constants and all conserved second
rank tensors built from the Riemann tensor and its derivatives are
proportional to the metric and the Ricci tensor. 

In this work, we shall show that there is a proper way to find universal
metrics in $d$-dimensions using curves in one less dimensions. This,
not so obvious, solution-generation that we shall lay out here, came
as a serendipitous surprise in our rather intense excursion to the
universal metric territory in the following works: we have shown that
the plane wave and spherical wave metrics, built on the anti-de Sitter
seeds, solve generic gravity theories \cite{Gullu-Gurses,Gurses-PRL,yeni_mak},
modulo the assumption that the Lagrangian is solely composed of the
curvature, covariant derivatives of the curvature and the metric tensor
in a Lorentz-invariant way (or the field equation is a covariantly
conserved two-tensor built from the metric). All of these solutions
are in the form of the Kerr-Schild--Kundt metrics 
\begin{equation}
g_{\mu\nu}=\bar{g}_{\mu\nu}+2V\lambda_{\mu}\lambda_{\nu},\label{eq:AdS-waveKS}
\end{equation}
where the \emph{seed} $\bar{g}_{\mu\nu}$ metrics are maximally symmetric,
whose explicit forms will be dictated by the curves that will generate
the solutions. The other ingredients of (\ref{eq:AdS-waveKS}) will
be discussed below. We first discuss the curves.

\vspace{0.3cm}

\section{Curves in Flat Spacetimes}

\noindent Let $z^{\mu}(\tau)$ define a smooth curve $C$ in ${\mathbb{R}}^{d}$,
with the metric $\eta_{\mu\nu}$. Here $\tau$ is the parameter of
the curve. From an arbitrary point $P(x^{\mu})$ not on the curve,
there are two null lines intersecting the curve at two points as shown
in the Figure.%
\footnote{Here, note that we take a generic curve such that it always has at
least one intersection with the null cone drawn from an arbitrary
point in the Minkowski spacetime.%
} 
\begin{figure}[h]
\includegraphics[width=0.4\textwidth]{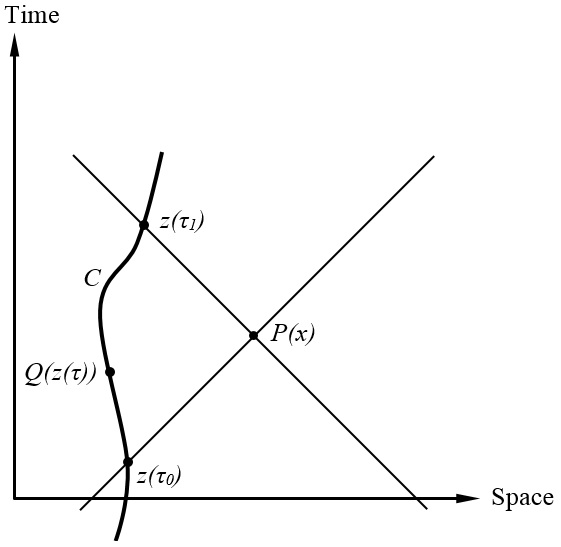} \protect\caption{Two null lines stretching from an arbitrary point $P\left(x\right)$
outside the curve meet the curve $C$ at the points corresponding
to the retarded and advanced times, that is $\tau_{0}$ and $\tau_{1}$,
respectively. $Q\left(z\left(\tau\right)\right)$ represents an arbitrary
point on the curve. }

\label{curve} 
\end{figure}

These intersection points are called the retarded ($\tau_{0}$) and
advanced ($\tau_{1}$) times \cite{vaidya,gur-sari-1}. Let $\Omega$
be the distance between the points $P(x^{\mu})$ and $Q(z^{\mu})$,
then since the spacetime is flat, it is simply given as 
\begin{equation}
\Omega^{2}=\eta_{\mu\nu}\,\big(x^{\mu}-z^{\mu}(\tau)\big)\,\big(x^{\nu}-z^{\nu}(\tau)\big),\label{denk1}
\end{equation}
which vanishes for the retarded and advanced times. There is a natural
null vector $\partial_{\mu}\tau_{0}$ that one can obtain by differentiating
$\Omega\left(\tau_{0}\right)=0$ with respect to $x_{\mu}$ as 
\begin{equation}
\ell_{\mu}\equiv\partial_{\mu}\,\tau_{0}=\frac{x_{\mu}-z_{\mu}(\tau_{0})}{R},\label{denk2}
\end{equation}
where $R$ is the retarded distance: $R\equiv\dot{z}^{\alpha}(\tau_{0})\,\big(x_{\alpha}-z_{\alpha}(\tau_{0})\big)$
with $\dot{z}^{\alpha}(\tau_{0})\equiv\partial_{\tau_{0}}z^{\alpha}(\tau_{0})$.
We have chosen to work with the retarded time $\tau_{0}$, but we
could equally have worked with the advanced time $\tau_{1}$ and the
ensuing results would not change. Moreover, in what follows, for notational
simplicity, we omit the subscript 0 from $\tau_{0}$ and use $\tau$
instead. Taking one more partial derivative of the null vector, one
has 
\begin{eqnarray}
\partial_{\nu}\ell_{\mu} & = & \frac{1}{R}\,\Bigl(\eta_{\mu\nu}-\dot{z}_{\mu}\,\ell_{\nu}-\dot{z}_{\nu}\,\ell_{\mu}-\big(A-\epsilon\big)\,\ell_{\mu}\,\ell_{\nu}\Bigr),\label{denklambda}
\end{eqnarray}
with $A\equiv\ddot{z}^{\mu}\,\left(x_{\mu}-z_{\mu}\right)$ and $\epsilon\equiv\dot{z}^{\mu}\,\dot{z}_{\mu}$,
($\epsilon=\pm1,0$), and the argument of $z^{\mu}$ and its derivatives
is always the retarded time.

\section{Universal Metrics}

The above has been a generic discussion of the curves in flat backgrounds.
Now comes the remarkable part of employing these curves to generate
solutions of generic gravity theories. Let us assume that the spacetime
metric is of the form (\ref{eq:AdS-waveKS}). Then, one can show that
the following relations hold for the metrics belonging to the Kerr-Schild--Kundt
class \cite{Gullu-Gurses,Gurses-PRL} 
\begin{equation}
\lambda^{\mu}\lambda_{\mu}=0,\qquad\nabla_{\mu}\lambda_{\nu}\equiv\xi_{(\mu}\lambda_{\nu)},\qquad\xi_{\mu}\lambda^{\mu}=0,\qquad\lambda^{\mu}\partial_{\mu}V=0.\label{eq:AdS-wave_prop}
\end{equation}
It is important to note that a new vector $\xi_{\mu}$ appears, besides
the two defining ingredients of the metric, the profile function $V$
and the vector $\lambda_{\mu}$. The first three relations describe
Kerr-Schild metrics belonging to the Kundt class and the last relation
is an \emph{assumption} which puts a further restriction on this class
of metrics. However, this last relation is crucial in proving the
universality of KSK metrics \cite{yeni_mak}. The covariant derivative
of $\lambda_{\mu}$ satisfies $\nabla_{\mu}\lambda_{\nu}=\bar{\nabla}_{\mu}\lambda_{\nu}$
where $\bar{\nabla}_{\mu}$ is the covariant derivative of the seed
metric. Then, for the AdS metric in the conformally flat coordinates
\begin{equation}
d\bar{s}^{2}=\frac{\ell^{2}}{z^{2}}\left(-dt^{2}+\sum_{m=1}^{d-2}\left(dx^{m}\right)^{2}+dz^{2}\right),\label{eq:AdS_can}
\end{equation}
$\nabla_{\mu}\lambda_{\nu}$ can be calculated as 
\begin{equation}
\nabla_{\mu}\lambda_{\nu}=\partial_{\mu}\lambda_{\nu}-\frac{1}{z}\eta_{\mu\nu}\lambda_{z}+\frac{1}{z}\left(\lambda_{\mu}\delta_{\nu}^{z}+\lambda_{\nu}\delta_{\mu}^{z}\right).
\end{equation}
On the other hand, for the dS seed metric in the conformally flat
coordinates 
\begin{equation}
d\bar{s}^{2}=\frac{\ell^{2}}{t^{2}}\left(-dt^{2}+\sum_{m=1}^{d-1}\left(dx^{m}\right)^{2}\right),\label{eq:dS_can}
\end{equation}
one has 
\begin{equation}
\nabla_{\mu}\lambda_{\nu}=\partial_{\mu}\lambda_{\nu}-\frac{1}{t}\eta_{\mu\nu}\lambda_{t}+\frac{1}{t}\left(\lambda_{\mu}\delta_{\nu}^{t}+\lambda_{\nu}\delta_{\mu}^{t}\right).
\end{equation}
By using these results and the defining expression $\nabla_{\mu}\lambda_{\nu}=\xi_{(\mu}\lambda_{\nu)}$
from (\ref{eq:AdS-wave_prop}), the partial derivative of $\lambda_{\mu}$
can be written, collectively for the AdS and dS, as 
\begin{equation}
\partial_{\nu}\,\lambda_{\mu}=a\,\eta_{\mu\nu}+\lambda_{\mu}\,\big(\frac{1}{2}\xi_{\nu}-\zeta_{\nu}\big)+\lambda_{\nu}\,\big(\frac{1}{2}\xi_{\mu}-\zeta_{\mu}\big)\label{defnofxi}
\end{equation}
where $a=\frac{\lambda_{z}}{z}$ , $\zeta_{\nu}=\frac{1}{z}\delta_{\nu}^{z}$
for the AdS seed \cite{Gullu-Gurses} and $a=-\frac{\lambda_{t}}{t}$,
$\zeta_{\nu}=\frac{1}{t}\delta_{\nu}^{t}$ for the dS seed.

The traceless-Ricci tensor, $S_{\mu\nu}\equiv R_{\mu\nu}-\frac{R}{d}g_{\mu\nu}$,
and the Weyl tensor, $C_{\mu\alpha\nu\beta}$, can be found, after
some tedious computation, as \cite{Gurses-PRL} 
\begin{equation}
S_{\mu\nu}=\rho\lambda_{\mu}\lambda_{\nu},~~C_{\mu\alpha\nu\beta}=4\lambda_{[\mu}\Omega_{\alpha][\beta}\lambda_{\nu]},\label{eq:Weyl_KSK}
\end{equation}
where the square brackets denote anti-symmetrization with a $1/2$
factor and the scalar function $\rho$ is given as 
\begin{equation}
\rho=-\left(\square+2\xi^{\mu}\partial_{\mu}+\frac{1}{2}\xi^{\mu}\xi_{\mu}-\frac{2\left(d-2\right)}{\ell^{2}}\right)V\equiv-{\cal Q}V.\label{rho}
\end{equation}
The second equality defines the operator ${\cal Q}$ which will play
a role in the field equations of the generic theory below. The symmetric
tensor $\Omega_{\alpha\beta}$, that appears in the Weyl tensor, can
be compactly written as 
\begin{equation}
\Omega_{\alpha\beta}\equiv-\left[\nabla_{\alpha}\partial_{\beta}+\xi_{(\alpha}\partial_{\beta)}+\frac{1}{2}\xi_{\alpha}\xi_{\beta}-\frac{1}{d-2}g_{\alpha\beta}\left({\cal Q}+\frac{2\left(d-2\right)}{\ell^{2}}\right)\right]V.
\end{equation}
For the seed metric, there are three possible choices whose explicit
forms are: 
\begin{eqnarray}
d\bar{s}^{2} & = & \frac{\ell^{2}}{\cos^{2}\theta}\,\Bigg(\frac{-du^{2}+2dudr}{r^{2}}+d\theta^{2}+\sin^{2}\theta\, d\omega^{2}\Bigg),\label{spherwave}\\
d\bar{s}^{2} & = & \frac{\ell^{2}}{z^{2}}\Big(du^{2}+2dudr+dx^{2}+\cdots+dz^{2}\Big),\label{planewave}\\
d\bar{s}^{2} & = & \frac{\ell^{2}}{\cosh^{2}\theta}\,\Bigg(\frac{du^{2}+2dudr}{r^{2}}+d\theta^{2}+\sinh^{2}\theta\, d\omega^{2}\Bigg),\label{hyperwave}
\end{eqnarray}
where $\ell$ is related to the cosmological constant and $d\omega^{2}$
is the metric of the $\left(d-3\right)$ unit-sphere. The first and
the second metrics are AdS metrics, while the third one is the dS
metric. 

Recently \cite{Gurses-PRL}, we have shown that the AdS-plane wave
and the $pp$-wave metrics in the Kerr-Schild form, and more generally
all Kerr-Schild--Kundt metrics are universal. The seed is the flat
Minkowski metric for the $pp$-waves, it is the AdS metric for the
AdS-plane and AdS-spherical waves, and it is the dS metric for the
dS-hyperbolic wave. Referring to \cite{Gurses-PRL,yeni_mak} for the
full proof, let us briefly recapitulate how this works.

Let the most general gravity theory be a $\left(2N+2\right)$-derivative
theory. As examples, for Einstein's gravity (and Einstein--Gauss-Bonnet
gravity) $N=0$, for quadratic and $f$(Riemann) theories $N=1$,
and for higher order theories $N\ge2$. We have shown that the equations
of the most general $\left(2N+2\right)$-derivative gravity theory
reduce, when evaluated for these metrics, to a rather compact form
\begin{equation}
eg_{\mu\nu}+\sum_{n=0}^{N}\, a_{n}\,\square^{n}\, S_{\mu\nu}=0,\label{denk0}
\end{equation}
where $e$ and $a_{n}$s are constants which are functions of the
parameters of the theory. Here, the constant $e$ determines the possible
effective cosmological constants in terms of the parameters of the
theory. After some algebraic manipulations, the traceless part of
(\ref{denk0}) reduces to a scalar equation of the metric function
$V$: 
\begin{equation}
\prod_{n=1}^{N}\,\big({\cal Q}-m_{n}^{2}\big)\,{\cal Q}\, V=0.\label{denk2}
\end{equation}
The generic solution is $V=V_{E}+\sum_{n=1}^{N}\, V_{n}$ where the
Einsteinian part ($V_{E}$) and the other (massive) parts satisfy
the following equations, respectively, 
\begin{eqnarray}
{\cal Q}V_{E}=0,\hskip1cm\big({\cal Q}-m_{n}^{2}\big)\, V_{n}=0,\label{denk3}
\end{eqnarray}
provided that all $m_{n}$'s are different and none is zero. If any
two or more $m_{n}$'s coincide and or equal to zero, then the second
equation in (\ref{denk3}) changes in the following way: let $r$
be the number (multiplicity) of $m_{n}$'s that are equal to $m_{r}$,
then the corresponding $V_{r}$ satisfies an irreducibly higher derivative
equation 
\begin{equation}
({\cal Q}-m_{r}^{2})^{r}\, V_{r}=0,\label{denk5}
\end{equation}
with new branches, so called log-solutions, appear. In that case,
the general solution becomes $V=V_{E}+V_{r}+\sum_{n=0}^{N-r}\, V_{n}$
and $V_{r}$ contains $\log^{r-1}$ terms.

Let us now get back to the issue of constructing these solutions from
the curves in flat space discussed in the previous section. The structural
similarity of the partial derivative of $\ell_{\mu}$ in (\ref{denklambda})
and the partial derivative of $\lambda_{\mu}$ in (\ref{defnofxi})
suggests the following procedure of generating Kerr-Schild--Kundt
class metrics: First, one takes the vectors $\ell_{\mu}$ and $\lambda_{\mu}$
in (\ref{denklambda}) and (\ref{defnofxi}) to be equal and derives
the corresponding vector $\xi_{\mu}$; and secondly, sets $\lambda^{\mu}\xi_{\mu}=0$
to satisfy the third condition in (\ref{eq:AdS-wave_prop}) and to
obtain the constraint on $z^{\mu}(\tau)$. The second step constrains
$z^{\mu}(\tau)$ curves to live in one less dimension.

Let us execute this procedure: when the seed metric is AdS as given
in (\ref{eq:AdS_can}), equating (\ref{denklambda}) and (\ref{defnofxi}),
one finds 
\begin{equation}
\xi_{\mu}=-\frac{2}{R}\,\left(\dot{z}_{\mu}+\frac{1}{2}\,(A-\epsilon)\,\lambda_{\mu}\right)+\frac{2}{z}\,\delta_{\mu}^{z}.\label{denk8}
\end{equation}
To satisfy $\lambda^{\mu}\,\xi_{\mu}=0$, we must have $\lambda_{z}=\frac{z}{R}$
and $z_{z}=0$.%
{} Hence, all these curves live in a $\left(d-1\right)$-dimensional
Minkowski spacetime. In this case, we have only timelike and null
curves. We can have spacelike curves, but the metrics generated by
these curves are equivalent to the metrics generated from timelike
curves via diffeomorphisms and possibly via complex transformations.
On the other hand, when the seed metric is the dS metric as given
in (\ref{eq:dS_can}), we find 
\begin{equation}
\xi_{\mu}=-\frac{2}{R}\,\left(\dot{z}_{\mu}+\frac{1}{2}\,(A-\epsilon)\,\lambda_{\mu}\right)+\frac{2}{t}\,\delta_{\mu}^{t}.\label{denk88}
\end{equation}
To satisfy $\lambda^{\mu}\,\xi_{\mu}=0$, we must have $\lambda_{t}=-\frac{t}{R}$
and $z_{t}=0$.%
{} Hence, the curve $C$ in this case lives in a $\left(d-1\right)$-dimensional
Euclidean space where we can have only spacelike curves. Let us turn
to some explicit examples.

\section{Explicit Examples}

We have infinitely many metrics characterized by the curves either
in the $\left(d-1\right)$-dimensional Minkowski space or the $\left(d-1\right)$-dimensional
Euclidean space. In the examples below, %
\begin{comment}
we consider only the %\textcolor[rgb]{1.00,0.00,0.00}{Kerr-Schild--Kundt%type of metrics when the generating curves are}straightlinesand
\end{comment}
{} for the sake of simplicity, we set $d=4$.

\noindent \vspace{0.3cm}
 \textbf{Example 1}: (Timelike case). Let $z_{\mu}=\tau\,\delta_{\mu}^{0}$,
then $\tau=t\pm r$. Choose $\tau=t-r$, then $R=-r$ and so one finds
\begin{equation}
\lambda_{\mu}=\Big(1,\frac{\vec{x}}{r}\Big),\hskip1cm\xi_{\mu}=\frac{2}{r}\,\Big(\delta_{\mu}^{0}-\frac{1}{2}\,\lambda_{\mu}\Big)+\frac{2}{z}\,\delta_{\mu}^{z}.
\end{equation}
This gives the AdS-spherical wave solution together with the profile
function $V$ solving the corresponding equations. For the explicit
form of $V$, see \cite{Gullu-Gurses}.

\noindent \vspace{0.3cm}
 \textbf{Example 2}: (Null case). If the curve is null, then $\epsilon=0$.
Let $z^{\mu}=\tau\, n^{\mu}$ where $\eta_{\mu\nu}\, n^{\mu}\, n^{\nu}=0$,
then we arrive at 
\begin{eqnarray}
\tau=\frac{x^{2}}{2n_{\mu}\, x^{\mu}},~~x^{2}=\eta_{\mu\nu}\, x^{\mu}\, x^{\nu},\hskip0.3cm\lambda^{\mu}=\frac{x^{\mu}-\tau\, n^{\mu}}{R},~~R=n_{\mu}\, x^{\mu},\, A=0,~~\lambda^{\mu}\, n_{\mu}=1.
\end{eqnarray}
Furthermore, one finds $\xi^{\mu}=-\frac{2}{R}\, n^{\,\mu}+\frac{2}{z}\,\delta_{z}^{\mu}$.
Choosing $n^{\mu}=(1,1,0,0)$ and performing a couple of coordinate
transformations, one obtains the AdS-plane wave metric 
\begin{equation}
ds^{2}=\frac{\ell^{2}}{\rho^{2}}\,\Big(2d\tau dv+d\sigma^{2}+d\rho^{2}\Big)+2V(\tau,\sigma,\rho)\, d\tau^{2}.\label{met1}
\end{equation}
\vspace{0.3cm}
 \textbf{Example 3}: (Spacelike Case). When the curve is spacelike,
the AdS seed is not allowed. However, de Sitter seed is possible,
\emph{i.e.}, $\bar{g}_{\mu\nu}=({\ell^{2}}/{t^{2}})\,\eta_{\mu\nu}$.
Then, for this case, the vector $\xi_{\mu}$ takes the form (\ref{denk88}).
Let $z^{\mu}=\tau\delta_{x}^{\mu}$, then we find $\tau=x\pm\sqrt{t^{2}-y^{2}-z^{2}}$
and $R=x-\tau=\mp\sqrt{t^{2}-y^{2}-z^{2}}$. Letting $r=\sqrt{t^{2}-y^{2}-z^{2}}$
and choosing the + sign, we get 
\begin{equation}
\lambda_{\mu}=\big(\frac{t}{r},1,-\frac{y}{r},-\frac{z}{r}\big).
\end{equation}
Letting $t=r\cosh(\theta)$, $y=r\,\sinh\theta\,\cos\phi$, $z=r\,\sinh\theta\,\sin\phi$,
the metric takes the form 
\begin{equation}
ds^{2}=\frac{\ell^{2}}{r^{2}\,\cosh^{2}\theta}\,\Bigl(du^{2}+2dudr+r^{2}\,(d\theta^{2}+\sinh^{2}\theta\, d\phi^{2})\Bigr)+2V(u,\theta,\phi)\, du^{2},
\end{equation}
for $\lambda_{\mu}=\delta_{\mu}^{0}$ and $\xi_{\mu}=\frac{1}{r}\,\delta_{\mu}^{0}+2\tanh(\theta)\,\delta_{\mu}^{2}$.
The above metric is the dS-hyperbolic wave metric given in (\ref{hyperwave})
which was noticed very recently \cite{dS-waves}.

%\noindent \vspace{0.3cm}%\textbf{4 .Concluding Remarks}:

\section{KSK Metrics in Robinson-Trautman Coordinates in Four Dimensions}

The metric form (\ref{eq:AdS-waveKS}) in the coordinates $\left(t,x,y,z\right)$
gives a very complicated expression for the operator ${\cal Q}$.
With this form, it is highly difficult to solve the equations in (\ref{denk3})
for the metric function $V$. In addition, we must also satisfy $\lambda^{\mu}\partial_{\mu}V=0$.
For this purpose, one should search for new coordinates where both
the metric and the operator ${\cal Q}$ take simpler forms. Two of
the new coordinates are (the natural coordinates) $\tau$ and $R$.
They are defined through $\Omega\left(\tau\right)=0$ and $R=\dot{z}^{\alpha}(\tau)\,\big(x_{\alpha}-z_{\alpha}(\tau)\big)$.
The coordinate transformation can be given as \cite{Newman_Unti}
\begin{equation}
x^{\mu}=R\lambda^{\mu}\left(\tau,\theta,\phi\right)+z^{\mu}\left(\tau\right),\qquad\mu=0,1,2,3.
\end{equation}
Here, the null vector $\lambda^{\mu}$ does not depend on the new
coordinate $R$ \cite{Newman_Unti}. In these new coordinates, we
have 
\begin{equation}
\partial_{R}V=\frac{\partial x^{\mu}}{\partial R}\partial_{\mu}V=\lambda^{\mu}\partial_{\mu}V=0.
\end{equation}
Hence, the metric function is independent of the new coordinate $R$.
Furthermore, in the new coordinates, $\lambda_{\mu}dx^{\mu}=d\tau$.
Hence, we have 
\begin{equation}
ds^{2}=d\bar{s}^{2}+2V\left(\tau,\theta,\phi\right)d\tau^{2},
\end{equation}
where $d\bar{s}^{2}$ is the background line element. The new form
of the metric in the new coordinates is called the Robinson-Trautman
(RT) metrics.

In four dimensions, to introduce the KSK metrics in the coordinates
of RT metrics, we first need the parametrizations of the two-dimensional
unit sphere and the two-dimensional unit hyperboloid. A parametrization
of the two-dimensional unit sphere, $(X^{1})^{2}+(X^{2})^{2}+(X^{3})^{2}=1$,
is given by the spherical coordinates
\begin{equation}
X^{1}=\sin\theta\,\sin\phi,\qquad X^{2}=\sin\theta\,\cos\phi,\qquad X^{3}=\cos\theta.\label{eq:S2_params}
\end{equation}

\noindent Similarly, the parametrization of a two-dimensional hyperboloid,
$-(Y^{0})^{2}+(Y^{1})^{2}+(Y^{2})^{2}=-1$, is given by
\begin{equation}
Y^{1}=\sinh\theta\,\sin\phi,\qquad Y^{2}=\sinh\theta\,\cos\phi,\qquad Y^{0}=\cosh\theta.\label{eq:H2_params}
\end{equation}

\subsection{The AdS Background }

Following \cite{Newman_Unti}, one can write the KSK metrics (\ref{eq:AdS-waveKS})
for AdS seed in the following form: 
\begin{equation}
ds^{2}=\frac{1}{f^{2}}\,\left(Hd\tau^{2}+2d\tau dr+\frac{r^{2}}{P^{2}}\,\left(d\theta^{2}+\sin^{2}\theta d\phi^{2}\right)\right)+2V\left(\tau,\theta,\phi\right)\, d\tau^{2},
\end{equation}
where the metric functions are

\noindent 
\begin{eqnarray}
H & = & \epsilon-2r\partial_{\tau}\log P,\qquad f=\frac{r}{\ell P}\,\cos\theta,\\
P & = & -\dot{z}^{0}(\tau)+\dot{z}^{1}(\tau)\, X^{1}+\dot{z}^{2}(\tau)\, X^{2},\\
\epsilon & = & -(\dot{z}^{0}(\tau))^{2}+(\dot{z}^{1}(\tau))^{2}+(\dot{z}^{2}(\tau))^{2}+(\dot{z}^{3}(\tau))^{2}.\label{unit4}
\end{eqnarray}
Here, $z^{\mu}=(z^{0}(\tau),z^{1}(\tau),z^{2}(\tau),z^{3}(\tau))$
is the parametrization of an arbitrary curve $C$ satisfying (\ref{unit4})
with $\epsilon=-1,0,1$, and $X^{i}$'s ($i=1,2,3)$ are defined in
(\ref{eq:S2_params}). As a result of the discussion above, the curve
$z\left(\tau\right)$ lives in one less dimension since $z^{3}(\tau)=0$.
The Ricci tensor takes the form 
\begin{equation}
R_{\mu\nu}=-\frac{3}{\ell^{2}}g_{\mu\nu}+\rho\lambda_{\mu}\,\lambda_{\nu},
\end{equation}
where $\lambda_{\mu}=\delta_{\mu}^{0}$ and the function $\rho$ has
the form given in (\ref{rho}). To calculate $\rho$ explicitly, one
needs to find $\xi_{\mu}$ from its defining relation 
\begin{equation}
\nabla_{\mu}\lambda_{\nu}=\bar{\nabla}_{\mu}\lambda_{\nu}=\xi_{(\mu}\lambda_{\nu)}.
\end{equation}
Here, $\bar{\nabla}$ is the covariant derivative of the AdS seed
which can be put in the form
\begin{equation}
d\bar{s}^{2}=\frac{1}{f^{2}}\left(Hd\tau^{2}+2d\tau dr\right)+\frac{\ell^{2}}{\cos^{2}\theta}\, g_{mn}dy^{m}dy^{n},
\end{equation}
with the metric of the two-dimensional unit sphere $g_{mn}$. Since
$\lambda_{\mu}=\delta_{\mu}^{0}$, one has 
\begin{equation}
\bar{\nabla}_{\mu}\lambda_{\nu}=-\bar{\Gamma}_{\mu\nu}^{\alpha}\lambda_{\alpha}=-\bar{\Gamma}_{\mu\nu}^{0},
\end{equation}
and from this relation, $\xi_{\mu}$can be calculated as 
\begin{equation}
\xi_{\mu}=2\partial_{\mu}\log f-2\delta_{\mu}^{r}\partial_{r}\log f+\frac{1}{2}\lambda_{\mu}f^{2}\partial_{r}\left(\frac{H}{f^{2}}\right).\label{eq:ksi_in_RT}
\end{equation}

\noindent Using this result in (\ref{rho}), after a long calculation,
the function $\rho$ is found to be
\begin{equation}
\rho=-{\cal Q}V=-\left(\bar{g}^{mn}\bar{\nabla}_{m}\partial_{n}+2\bar{g}^{mn}\partial_{m}\log f\partial_{n}+2\bar{g}^{mn}\partial_{m}\log f\partial_{n}\log f-\frac{4}{\ell^{2}}\right)V,\label{eq:ro2}
\end{equation}
where $\bar{g}_{mn}\equiv\frac{\ell^{2}}{\cos^{2}\theta}\, g_{mn}$.
Clearly, the operator ${\cal Q}$ contains derivatives only with respect
to the angular coordinates and can be found once one has the explicit
form of the curve is given. Then, one can solve the massless and massive
wave equations given in (\ref{denk3}).

\subsection{The dS Background}

Since the dS case follows similar to the AdS case (albeit with subtle
differences) above, without much ado let us give the metric and the
relevant results. First, the KSK metrics (\ref{eq:AdS-waveKS}) for
dS seed becomes 
\begin{equation}
ds^{2}=\frac{1}{f^{2}}\,\left(Hd\tau^{2}+2d\tau dr+\frac{r^{2}}{P^{2}}\,\left(d\theta^{2}+\sinh^{2}\theta\, d\phi^{2}\right)\right)+2V\left(\tau,\theta,\phi\right)\, d\tau^{2},
\end{equation}
where the metric functions are

\noindent 
\begin{eqnarray}
H & = & 1-2r\partial_{\tau}\log P,\qquad f=\frac{r}{\ell P}\,\cosh\theta,\\
P & = & \dot{z}^{1}(\tau)\, Y^{1}+\dot{z}^{2}(\tau)\, Y^{2}+\dot{z}^{3}(\tau),\label{hypf}\\
1 & = & (\dot{z}^{1}(\tau))^{2}+(\dot{z}^{2}(\tau))^{2}+(\dot{z}^{3}(\tau))^{2}.\label{unit5}
\end{eqnarray}
Here, $Y^{i}$'s ($i=1,2)$ are defined in (\ref{eq:H2_params}) and
$z^{0}(\tau)=0$. The Ricci tensor takes the form 
\begin{equation}
R_{\mu\nu}=\frac{3}{\ell^{2}}g_{\mu\nu}+\rho\lambda_{\mu}\,\lambda_{\nu},
\end{equation}
where $\lambda_{\mu}=\delta_{\mu}^{0}$. The dS seed can be put in
the form
\begin{equation}
d\bar{s}^{2}=\frac{1}{f^{2}}\left(Hd\tau^{2}+2d\tau dr\right)+\frac{\ell^{2}}{\cosh^{2}\theta}\, g_{mn}dy^{m}dy^{n},
\end{equation}
with the metric of the two-dimensional unit hyperboloid $g_{mn}$.
Again, to find the function $\rho$, one needs to find $\xi_{\mu}$which
takes the same form (\ref{eq:ksi_in_RT}). Then, the function $\rho$
in (\ref{rho}) becomes 
\begin{equation}
\rho=-\left(\bar{g}^{mn}\bar{\nabla}_{m}\partial_{n}+2\bar{g}^{mn}\partial_{m}\log f\partial_{n}+2\bar{g}^{mn}\partial_{m}\log f\partial_{n}\log f+\frac{4}{\ell^{2}}\right)V,
\end{equation}
where $\bar{g}_{mn}\equiv\frac{\ell^{2}}{\cosh^{2}\theta}\, g_{mn}$.

\section{Conclusion}

We have given a way of constructing the Kerr-Schild--Kundt type of
metrics which we have shown previously to be universal metrics of
generic purely metric-based theories of gravity. It is highly interesting
that we have three families of curves generating all these universal
metrics. When the seed metric is the AdS spacetime, we have two families
corresponding to timelike and null curves. They generate the AdS-plane
wave and AdS-spherical wave families. For the dS seed metric, only
the spacelike curves generate the dS-hyperbolic wave family. Hence,
we obtain, in principle, an infinite number of Kerr-Schild--Kundt
type of metrics where the AdS-plane wave, AdS-spherical wave \cite{Gullu-Gurses},
and dS-hyperbolic wave metrics \cite{dS-waves} correspond to the
straight lines and the rest of the family offer an exciting new territory
of investigation. Using the Robinson-Trautman coordinates, we recast
the KSK metrics in a convenient form which is suitable for studying
explicit solutions.

\section*{Acknowledgment}

This work is partially supported by TUBITAK. M.~G. and B.~T. are
supported by the TUBITAK grant 113F155. T.~C.~S. is supported by
the Science Academy's Young Scientist Program (BAGEP 2015).

\end{document}